\pdfoutput=1

\documentclass[11pt]{article}

\usepackage[preprint]{acl}

\usepackage{times}
\usepackage{latexsym}

\usepackage[T1]{fontenc}

\usepackage[utf8]{inputenc}

\usepackage{microtype}

\usepackage{inconsolata}

\usepackage{caption}
\usepackage{graphicx}
\usepackage{float} 
\usepackage{subcaption}
\usepackage{amsmath}
\usepackage{amssymb}
\usepackage{hyperref}
\usepackage{url}
\usepackage{hyperref}
\usepackage{booktabs}
\usepackage{enumitem}
\usepackage{multirow}
\usepackage[linesnumbered,ruled]{algorithm2e}
\SetKwComment{Comment}{/* }{ */}
\RestyleAlgo{ruled}

\newcommand{\M}{SVPT}

\title{Self-Supervised Singing Voice Pre-Training towards Speech-to-Singing Conversion}

\author{
Ruiqi Li, 
Rongjie Huang,
Yongqi Wang, 
Zhiqing Hong,  
Zhou Zhao\thanks{Corresponding author}
\\
Zhejiang University \\
\texttt{\{ruiqili,rongjiehuang,zhaozhou\}@zju.edu.cn}
}

\begin{document}
\maketitle
\begin{abstract}
Speech-to-singing voice conversion (STS) task always suffers from data scarcity, because it requires paired speech and singing data. Compounding this issue are the challenges of content-pitch alignment and the suboptimal quality of generated outputs, presenting significant hurdles in STS research. This paper presents SVPT, an STS approach boosted by a self-supervised singing voice pre-training model.
We leverage spoken language model techniques to tackle the rhythm alignment problem and the in-context learning capability to achieve zero-shot conversion. We adopt discrete-unit random resampling and pitch corruption strategies, enabling training with unpaired singing data and thus mitigating the issue of data scarcity. SVPT also serves as an effective backbone for singing voice synthesis (SVS), offering insights into scaling up SVS models. Experimental results indicate that SVPT delivers notable improvements in both STS and SVS endeavors. Audio samples are available at \href{https://speech2sing.github.io}{speech2sing.github.io}.

\end{abstract}

\section{Introduction}

A speech-to-singing voice conversion (STS) system \cite{cen2012template, parekh2020speech, li2023alignsts} transforms the semantic content of a speech signal into a paired singing signal, conditioned on any form of pitch information, such as fine-grained F0 sequence \cite{saitou2007speech}. Beyond being part of the development of music entertainment, works on STS provide insightful perspectives on bridging the gap between well-developed speech-language models and rudimentary singing voice modeling.

    
Although research in STS has achieved notable success recently, it continues to encounter several challenges. The major concern is the scarcity of paired speech-singing voice data. Previous works rely on pre-made paired datasets \cite{duan2013nus, sharma2021nhss}, which have a much smaller quantity than singing voice corpora. The scarcity of data also makes STS models difficult to generalize or scale up. In addition, the non-autoregressive design in previous works exacerbates the misalignment issue. Although \citet{li2023alignsts} tries to solve this problem by utilizing a roughly monotonic cross-attention mechanism, the attention map has a chance to degrade and corrupt when facing complex, long sentences. 

\begin{figure}[htbp]
    \centering
    \includegraphics[width=0.48\textwidth]{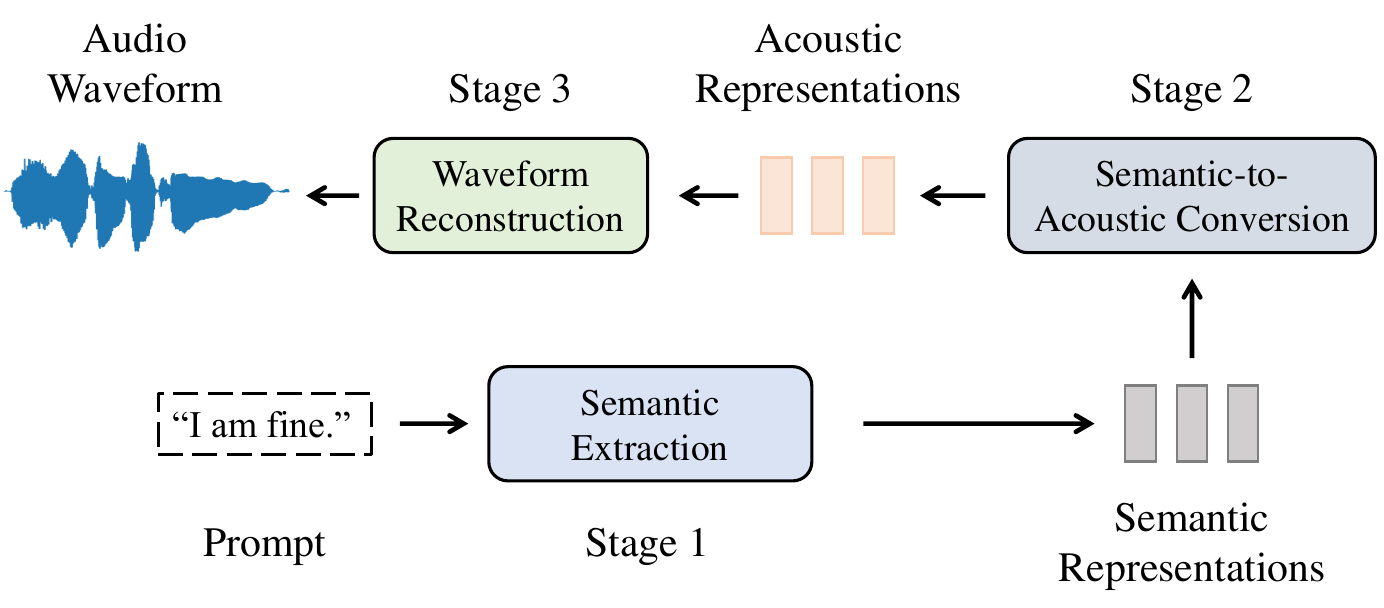}
    \caption{Speech/Singing synthesis paradigm.}
    \label{fig:tts-scheme}
\end{figure}

Recent breakthroughs in large-scale generative spoken language models and audio discretization methods have opened up new frontiers in speech generation \cite{lee2022hierspeech, borsos2023audiolm, wang2023neural}. To lighten the burden of directly modeling the target acoustic space, researchers split the models into two stages and map the prompts into an intermediate semantic latent space first, as illustrated in \autoref{fig:tts-scheme}. These achievements, however, have not significantly benefited the field of singing voice synthesis (SVS) \cite{Liu_Li_Ren_Chen_Zhao_2022, 9747664}. Because, unlike speech, the highly dynamic nature of singing voice and the scarcity of annotated data make it a challenge for language models (LMs). Either the frequency domain (pitch information) or temporal domain (rhythm component, or phoneme durations) of singing voices is highly unstable. 
Luckily, these problems and the challenges of STS mentioned before have a chance to solve each other. 
STS models do not need textual annotations, while the in-context learning capability of spoken language models allows for an effective solution to the misalignment issue. 
As shown in \autoref{fig:tts-scheme}, stage 2, translating high-level semantic representations into low-level acoustic representations, requires no annotations or transcriptions. 
Similarly, \citet{li2023alignsts} disentangles the input speech signal into corrupted semantic-related content representations and transforms them into Mel-spectrograms, conditioned on pitch information. 
Therefore, stage 2 in a spoken language model is naturally suited to STS tasks.
We claim that with proper regulations and enhancements, spoken language models are self-supervised singing voice learners, which make use of abundant unannotated data and bridge the gap between speech and singing voice. 

In this paper, we propose \M, an STS approach boosted by a self-supervised \textbf{S}inging \textbf{V}oice \textbf{P}re-\textbf{T}raining model, which further improves singing voice synthesis. 
We construct a multi-scale Transformer with a decoder-only architecture \cite{yang2023uniaudio}, specializing in long-sequence modeling like discrete audio tokens. Special regulations and perturbations, such as discrete-unit random resampling and expanded-range reference prompting, are introduced to generalize and stabilize the model. 
Essentially, the proposed model translates the corrupted semantic tokens into the target acoustic tokens autoregressively, conditioned on pitch information and sampled reference prompt tokens. 
The benefits are three-fold:
(a) the models extracting the semantic and acoustic tokens are pre-trained and finetuned in a self-supervised fashion, making this procedure fully annotation-free; 
(b) the semantic and acoustic tokens can be derived from singing voice samples solely, where the former are corrupted and disentangled to have a closer distribution distance to speech signals; 
(c) with an external text-to-semantic translator, the STS model can be upgraded to a high-quality zero-shot singing voice synthesizer. 
Our contributions are summarized as follows:
\begin{itemize}[leftmargin=*]
    \setlength{\itemsep}{0pt}
    \item We propose SVPT, the first STS approach boosted by a self-supervised singing voice pre-training model. 
    \item We introduce special regulations and enhancements to generalize the discrete spoken language model to singing voices, which are more dynamic in frequency or temporal domains. 
    \item We explore the connection between STS and SVS tasks.
    SVPT can be generalized into a zero-shot SVS model with an external text-to-semantic transformer.
\end{itemize}

\section{Related Works}

\subsection{Speech-to-Singing Voice Conversion}

Three major approaches are developed: a) Model-based approaches rely on phone-score synchronization information and manual alignment with multiple artificial control models \cite{saitou2004analysis,saitou2007speech}. 
b) Template-based approaches \cite{cen2012template,vijayan2017dual,vijayan2018analysis} require an available high-quality reference vocal input with dynamic time warping (DTW)-based alignment. 
c) Style transfer-based methods \cite{parekh2020speech} view STS as a style-transfer problem and apply deep-learning approaches, not requiring external synchronization or a high-quality template. \citet{parekh2020speech} designs a CNN-based network and re-aligns the signal by simply stretching the speech to the target length. By leveraging boundary-equilibrium GAN (BEGAN) \cite{berthelot2017began}, \citet{wu2020speech} expands their previous work to improve the synthesis quality. \citet{li2023alignsts} introduces the rhythm component of spoken voices and uses a cross-attention modality fusion to improve the alignment. 

\subsection{Generative Spoken Language Models}

Generative spoken language model (GSLM) \cite{lakhotia2021generative} is originally proposed to tackle the speech generation problem in a textless setting. AudioLM \cite{borsos2023audiolm} uses semantic tokens from a pre-trained w2v-BERT model \cite{chung2021w2v} and acoustic codecs \cite{zeghidour2021soundstream} to construct a coarse-to-fine decoder-only architecture to model audios. MusicLM \cite{agostinelli2023musiclm} continues the work and focus on the field of music generation, leveraging a text-music joint representation \cite{huang2022mulan}. VALL-E \cite{wang2023neural} uses a similar paradigm to build a zero-shot text-to-speech (TTS) model, with seven additional non-autoregressive (NAR) decoders to reconstruct fine-grained acoustic codes. Audio tokenizers \cite{zeghidour2021soundstream, défossez2022high} laid a good foundation for audio codec LMs. SPEAR-TTS \cite{kharitonov2023speak} interprets the first two stages in \autoref{fig:tts-scheme} as reading and speaking respectively, alleviating the demand for annotated data. Regarding the generation patterns of audio tokens, MusicGen \cite{copet2024simple} investigates different codebook interleaving patterns and proposes a one-stage music generation model. 
UniAudio \cite{yang2023uniaudio} and AudioBox \cite{vyas2023audiobox} further generalize the paradigm to unified audio generation. 

\subsection{Singing Voice Synthesis}

Recently, remarkable progress has emerged in the field of SVS. \citet{chen2020hifisinger} and \citet{zhang2022wesinger} adopt GAN-based networks for high-quality synthesis. DiffSinger \cite{Liu_Li_Ren_Chen_Zhao_2022} designs a shallow diffusion mechanism to solve the problem of over-smoothness in the general TTS field. Inspired by VITS \cite{kim2021vits}, VISinger \cite{9747664} constructs an end-to-end architecture. For singer generalization, NaturalSpeech 2 \cite{shen2023naturalspeech} and StyleSinger \cite{zhang2023stylesinger} use a reference voice clip for timbre and style extraction. 


\section{Method}


\subsection{Overview}

As stated, a common audio generation process can be summarized in 3 stages, as shown in \autoref{fig:tts-scheme}. 
In the context of STS, the prompts are input speech signals. However, the speech input has the same semantic content as the target singing voice, making the demand for speech data in the training stage no longer necessary. 
Inspired by \cite{li2023alignsts}, we use several information perturbation operations as enhancements, corrupting the singing representations without altering the semantic content. In stage 2, the pitch information is involved to guide the rhythm and pitch reconstruction, while a reference voice is appended to guide timbre reconstruction. Finally, we leverage a unit-based vocoder to synthesize the high-fidelity waveforms from acoustic codes. The whole process, including the extraction of semantic and acoustic tokens, is self-supervised with unannotated singing voice data.

Although our model requires no text transcription, it has the potential to be an SVS model. We adopt an external text-to-semantic transformer trained with speech/singing voice data, which is of a relatively smaller amount, comparing the unannotated data used in stage 2. 

\subsection{Voice Discretization}

Inspired by previous works, we adopt a coarse-to-fine hierarchy with discrete-unit audio modeling. One important benefit of this design in an STS model is that the intermediate semantic space covers both speech and singing distributions. In this section, we briefly introduce the voice discretization strategies.

\textbf{Semantic tokens} \ The semantic modeling stage generates high-level, intermediate representations with rich linguistic information. Under this requirement, we leverage XLSR-53 \cite{conneau2020unsupervised}, a wav2vec 2.0 \cite{baevski2020wav2vec} model pre-trained on 56k hours of speech in 53 languages. 
We extract the features from the 12th layer, which are the most relevant to pronunciation features \citet{singla2022audio}. 
We also extract the features from the 18th layer for comparison, which is more focused on semantic information. 
Finally, we apply a k-means algorithm to the features to cluster $K_1$ centroids, the indices of which are used as discrete units. Formally, given a singing voice signal $\boldsymbol{y}$, we extract a unit sequence $\boldsymbol{s} = \mathcal{F}(\boldsymbol{y}) = [s_1, s_2, ..., s_T]$, where $\mathcal{F}$ is the combination function of the operations above, and $s_t \in \{0, 1, ..., K_1 - 1\}, \forall t \in [1, T]$. 

\vspace{1pt}

\textbf{Acoustic tokens} \ We use SoundStream, an audio codec model, to produce tokenized acoustic features. The codec model applies residual vector quantization with $N_q = 8$ codebooks. Therefore, given a sample $\boldsymbol{y}$, we produce a vector sequence $\boldsymbol{A} = 
\mathcal{G}(\boldsymbol{y}) = [\boldsymbol{a}_1, \boldsymbol{a}_2, ..., \boldsymbol{a}_T]$, where $\mathcal{G}$ is the encodec operation and $\boldsymbol{a}_t = [a_t^{1}, a_t^{2}, ..., a_t^{N_q}]$, being a vector contains the codes from all the codebook at timestep $t \in [1, T]$, and $a_t^{\tau} \in \{0, 1, ..., K_2 - 1\}, \forall \tau \in [1, N_q]$. $K_2$ is the codebook length of all the codebooks. To lighten the burden of autoregressive generation, we use the stacked codes from the first 3 codebooks as the discrete acoustic features. To recover the loss, we leverage an additional unit-based vocoder, BigVGAN, to reconstruct high-fidelity waveforms \cite{lee2022bigvgan}.

\vspace{1pt}

\textbf{Pitch and reference tokens} \ The generation of acoustic tokens is conditioned not only on semantic tokens, but also on pitch information and reference tokens. We extract the fundamental frequency contour F0 as pitch information. For discretization, we simply round the frequency numbers to integers, creating a pitch token sequence $\boldsymbol{p} = [p_1, p_2, ..., p_T]$, where $p_t \in [f_{\text{min}}, f_{\text{max}}], \forall t \in [1, T]$. 
The reference tokens are similar to the acoustic tokens, except that we only use the codes from the first codebook. 

\subsection{Information Perturbation}


Compared with speech corpus, singing voice data is different in two ways: a) singing voice corpus has a significantly smaller quantity, and b) either the frequency domain (pitch) or temporal domain (phoneme durations, or rhythm) of singing voices is highly dynamic. These features make self-supervised training extremely unstable and vulnerable to over-fitting. To ensure that the semantic tokens consist of only semantic-related information, disentanglement is required. 


\subsubsection{Pitch and Timbre Corruption}

Since the pitch component of a singing voice sample is correlated with the speaker identity, we corrupt the pitch and timbre features at the same time. Inspired by \citet{choi2021neural}, we use a chain of three functions to generate a corrupted waveform $\boldsymbol{\overline{y}} = F_p(\boldsymbol{y}) = fs(pr(peq(\boldsymbol{y})))$, where $fs(\cdot)$ is formant shifting function, $pr(\cdot)$ is pitch randomization, and $peq(\cdot)$ is parametric equalizer. By doing so, the corrupted sample $\boldsymbol{y}$ contains only semantic-related information to the fullest extent. More details and hyperparameters are listed in \autoref{sec:info}.

It is worth mentioning that training the overall model while perturbing the waveforms and extracting semantic tokens on the fly consumes a great amount of computing resources, which is unrealistic for an academic laboratory. Therefore, we randomly pre-perturb the dataset $N_r$ times, resulting in a $N_r\times$ larger semantic corpus. Experiments show that $N_r = 20$ reaches satisfactory performance. 

\subsubsection{Rhythm Corruption}

Unlike speech voices, the rhythm information of singing voices is crucial and has the potential to be leaked in semantic representations. More importantly, the rhythm component differs between speech and singing voices, creating a non-negligible domain shift. Suppose there is a perturbation operation that isolates the semantic information from singing voices and makes it compatible with speech voices at the same time, we can pre-train the model on unannotated singing data and generalize it to speech inputs. 

Inspired by \citet{chan2022speechsplit2}, we adopt temporal random resampling operation to tackle the challenge. The original random resampling algorithm retains source length, which is not necessary in STS. To accelerate the computation without loss of effectiveness, we design a pseudo random resampling algorithm $\boldsymbol{\widetilde{x}} = PRR(\boldsymbol{x})$, shown in \autoref{alg:prr}. The algorithm accepts a hyperparameter, average segmentation length $l_{r}$, to cut the signal into several segments. Each segment is randomly scaled up or down temporally and finally concatenated together. Therefore, the original distinguishable rhythm information is disrupted and removed. 


\begin{figure}[htbp]
    \centering
    \includegraphics[width=0.48\textwidth]{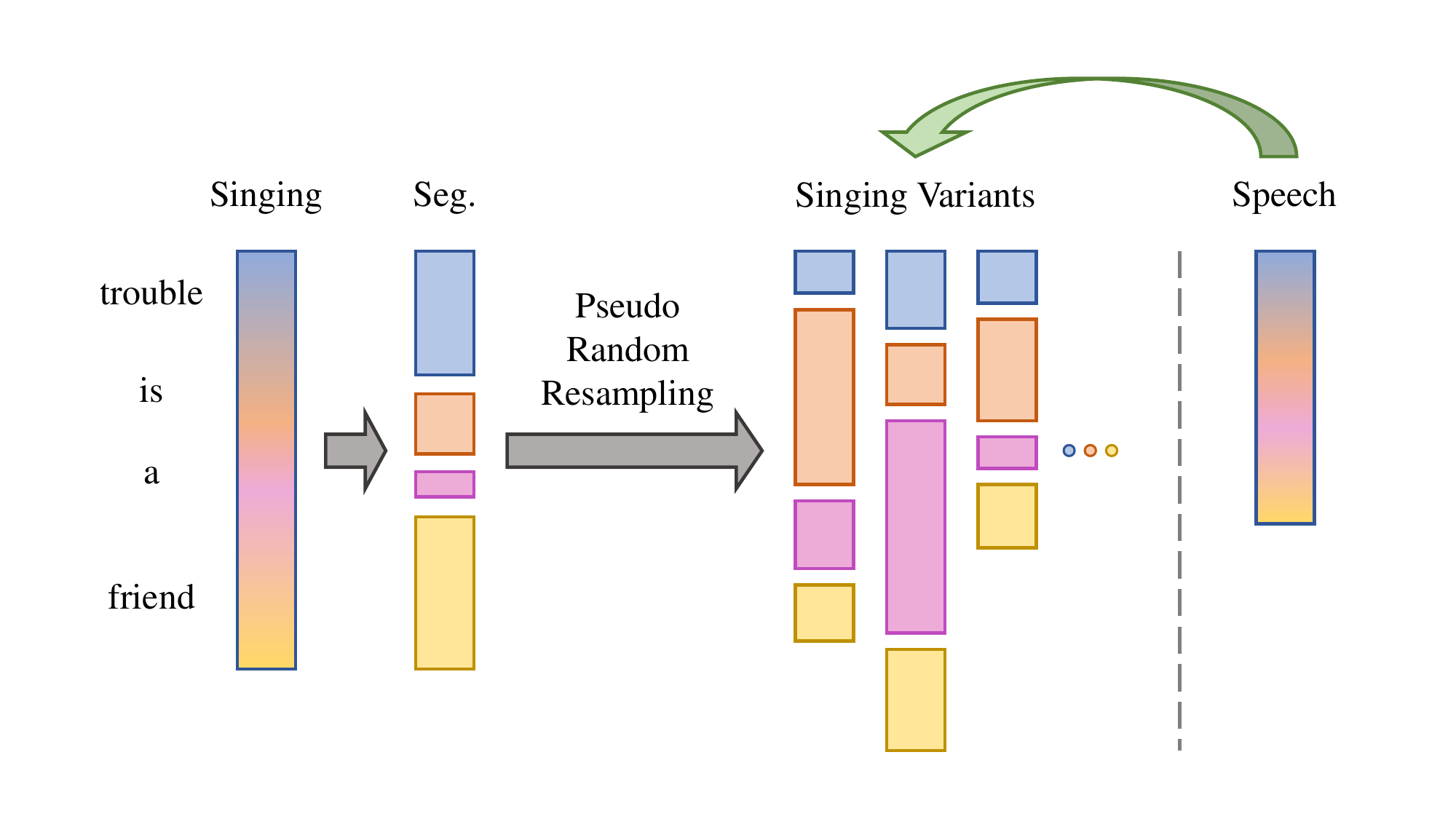}
    \caption{Pseudo random resampling.}
    \label{fig:rr}
\end{figure}

Hypothetically, the random resampling operation $PRR(\cdot)$ re-aligns the durations of phonemes, or the rhythm component. Although the phoneme durations of singing voices change drastically compared to speech, it has a small probability of re-aligning the signal that is close enough to the speech signal. In other words, if the model sees enough re-alignment results of corrupted singing signals, the speech input in the inference stage will just be treated as another re-alignment, as illustrated in \autoref{fig:rr}. 

The hyperparameter $l_{r}$ is an important factor of random resampling effectiveness. If $l_{r}$ is too large, more than one phoneme could be grouped in one segment, attenuating the actual effect of resampling. If $l_{r}$ is too small, distortion may occur due to round-off error. We set $l_{r}$ to 0.4 seconds, and a little discussion can be found in \autoref{sec:lr}.

Normally, $PRR(\cdot)$ is performed on continuous signals like waveforms or spectrograms, implemented by linear interpolation. However, the issue of computational overhead mentioned before also exists. Therefore, we adopt two strategies:
\begin{itemize}[leftmargin=*]
    \setlength{\topsep}{0pt}
    \setlength{\itemsep}{0pt}
    \item \textbf{Continuous resampling $PRR_C(\cdot)$.} We still adopt random resampling on waveforms with linear interpolation, but for $N_r$ times in advance, similar to pitch pre-perturbation. Practically, we perform both pitch and rhythm perturbations $N_r$ times in advance, resulting in a singing variant corpus $N_r$ times the original size. 
    \item \textbf{Discrete resampling $PRR_D(\cdot)$.} Instead of the continuous operation, we perform discrete random resampling on the clustered semantic tokens implemented by nearest neighbor interpolation. The disadvantages are obvious: the semantic content is inherently continuous, while discrete operations can introduce abrupt transitions and artifacts. As a result, the semantic tokens derived from a resampled waveform might not match those obtained directly from the same signal, even if the resampling path remains unchanged. Despite this, we believe that the additional noise introduced by this drawback just involves more perturbations, forcing the model to focus on the denoised content. Furthermore, this method allows for on-the-fly resampling, creating a theoretically infinite variant corpus. 
\end{itemize}

It is worth mentioning that both strategies are considered discrete-unit random resampling, since the results are discrete tokens. 




\subsection{Expanded-range Reference Prompting}

The objective of reference prompting is to modulate the generated vocal traits via a specific reference prompt sequence \cite{borsos2023audiolm}, enabling language models to produce novel voices in a zero-shot fashion. Prior approaches choose two distinct speech windows from each training sample—one as the reference prompt and the other as the intended output, as described in \cite{kharitonov2023speak}—this method's efficiency is somewhat diminished with singing voice data due to its scarcity, leading to potential model overfitting. To mitigate this, we broaden the selection of reference samples. Initially, our experiments involved selecting a random sample from the same singer as the reference. Nevertheless, variations in style and timbre within the same singer's different songs prompted us to refine our approach: we now select a random sample from the same song within a certain range to ensure consistency in vocal characteristics. Specifically, we only sample segments within the range from the 5 preceding to the 5 following ones in the original order of the song as references. 

\subsection{Architecture and Training}

\begin{figure*}[htbp]
\centering
\includegraphics[width=\textwidth]{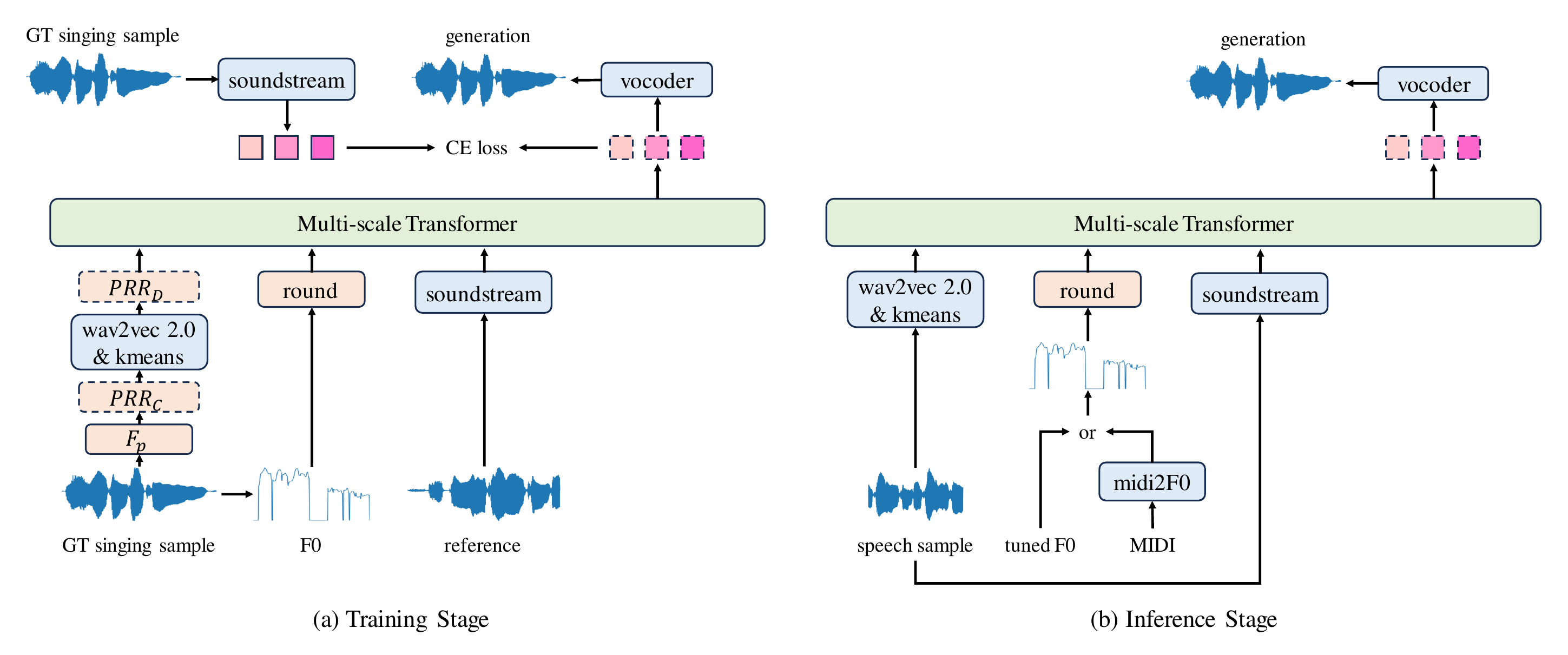}
\setlength{\abovecaptionskip}{-0.4cm}
\setlength{\belowcaptionskip}{-0.4cm}
\caption{Model architecture and the training/inference stage. }
\label{fig:main-model}
\end{figure*}


\subsubsection{Multi-scale Transformer Architecture}

To address the challenge of processing long-sequence audio tokens, we introduce a multi-scale Transformer based on a decoder-only framework, inspired by \citet{yang2023uniaudio}. This multi-scale model consists of a global and a local Transformer, respectively. Initially, the sequence is segmented into patches of equal length $P$, with the embeddings of tokens within each patch being merged to reduce the sequence's temporal resolution. The global layer, denoted as $\theta_{\text{G}}$, processes this compressed sequence to autoregressively produce a sequence of hidden features $\boldsymbol{h}$. Concurrently, the local layer, $\theta_{\text{L}}$, utilizes the aggregated global hidden feature pertinent to its patch and the original tokens of the patch to model the local context. Specifically, we arrange acoustic codes side by side, where every trio of consecutive codes from distinct codebooks delineates a single patch (equating one patch to one frame). To accommodate the structure for semantic, pitch, and reference tokens, we triple each token, ensuring compatibility with the framework's requisites.
Therefore, the patch size $P=3$ and the local Transformer can be formulated as:

\begin{align}
    & p(\boldsymbol{a}_t | \hat{h_t}, rp_P(\widetilde{s}_t), rp_P(p_t), rp_P(r_t); \theta_{\text{L}}) = \\
    & \prod_{\tau=1}^{P} p(a_t^{\tau} | \boldsymbol{a}_{t}^{<\tau}, \hat{h_t}, rp_P(\widetilde{s}_t), rp_P(p_t), rp_P(r_t); \theta_{\text{L}}) \nonumber
\end{align}
where $rp_P(\cdot)$ repeats the element $P$ times and concatenates them temporally. $\boldsymbol{a}_t$ indicates the acoustic tokens from $P$ codebooks at timestep $t$, while $\widetilde{s}_t$, $p_t$, and $r_t$ indicate the corrupted semantic token, the pitch token, and the reference token. $\hat{h_t}$ is the hidden output at timestep $t$ from the global Transformer:
\begin{align}
    p(\boldsymbol{\hat{h}} | \boldsymbol{h}, \boldsymbol{\widetilde{s}}, \boldsymbol{p}, \boldsymbol{r}; & \theta_{\text{G}}) = \\ \nonumber
    & \prod_{t=1}^{T} p(\hat{h}_t | \boldsymbol{\hat{h}}_{<t}, \boldsymbol{h}_{<t}, \boldsymbol{\widetilde{s}}, \boldsymbol{p}, \boldsymbol{r}; \theta_{\text{G}})
\end{align}
where $\boldsymbol{h}$ is the downsampled sequence by concatenating the embeddings channel-wise within one patch: $h_t = \text{concat}([\text{embed}(a_t^{1}), ..., \text{embed}(a_t^{P})])$. By maximizing the probability $\prod_{t=1}^{T} p(\boldsymbol{a}_t | \boldsymbol{a}_{<t}, *)$, the model approximates the distribution of acoustic representations and generates acoustic codes autoregressively.

\subsubsection{Training and Inference}

As stated before, during the training stage we use unannotated singing data to train the transformers. The corrupted semantic tokens, the pitch tokens, and the reference tokens are concatenated temporally with special separator tokens such as \texttt{<semantic\_start>}, \texttt{<acoustic\_end>}, \texttt{<pitch\_start>}, etc. The reference inputs are also singing voices, sampled in the same song. The training object is the cross entropy loss. As for inference, the model takes unseen speech samples to perform zero-shot STS. The speech sample is also considered as the reference to extract style and timbre. The pitch information can be provided by hand-tuned or auto-tuned F0 contour, or predicted by a pre-trained midi2F0 model, which takes MIDI notes as input. 

\subsection{A Leap towards Singing Voice Synthesis}

In addition to its capabilities for STS, the pre-trained model serves as an effective foundation for SVS tasks. This model, designed to extract only semantic information from the input, can be enhanced with an auxiliary text-to-semantic translator to facilitate text-based synthesis. We develop this translator using a smaller annotated speech and singing dataset, focusing solely on semantic extraction. For this purpose, a 12-layer transformer is employed to generate semantic tokens from phonemes in an autoregressive manner. This auxiliary component enables the transformation of the STS model into a high-fidelity SVS model, expanding its capacity to leverage a larger amount of unannotated data efficiently. 

\section{Experiments}

\subsection{Data} \label{sec:exp-data}

\begin{table}[ht]
\centering
\setlength{\belowcaptionskip}{-0.4cm}
\begin{tabular*}{\hsize}{l|c|c}
\toprule
\bf Training Datasets         & \bf L    & \bf Hours  \\
\midrule
Opencpop \cite{wang2022opencpop}        & ZH         & 5.3       \\
Opensinger \cite{huang2021multi}        & ZH         & 83.5       \\
M4Singer \cite{zhang2022m4singer}       & ZH         & 29.8       \\
internal \#1                            & ZH         & 22.1       \\
internal \#2                            & ZH         & 21.0       \\
CSD \cite{choi2020children}             & EN          & 1.9       \\
musdb18hq \cite{MUSDB18HQ}             & EN          & 5.7       \\
VocalSet \cite{wilkins2018vocalset}     & EN          & 7.6        \\
internal \#3                            & EN          & 3.0        \\
\bottomrule
\end{tabular*}
\caption{\label{tab:train-data}
Information of training datasets.
}
\end{table}

We construct a bilingual singing voice training corpus spanning 180 hours, with dataset specifics outlined in \autoref{tab:train-data}. This includes 161.7 hours of Mandarin and 18.2 hours of English audio. To enrich our dataset further, we augment it with an additional 10 hours of speech data each from AISHELL-3 \cite{shi2020aishell} and LibriSpeech \cite{panayotov2015librispeech}, bringing our total training corpus to 200 hours. For validation and testing, we randomly allocate two segments of 4 hours each. In terms of baseline training, we utilize two STS datasets: the NHSS \cite{sharma2021nhss} and PopBuTFy \cite{liu2022learning} databases. NHSS provides 7.0 hours of English data, split between 2.3 hours of speech and 4.7 hours of singing, while PopBuTFy offers 18 hours, divided into 8 hours of speech and 10 hours of singing.

For zero-shot STS inference, we collect a bilingual corpus of paired speech and singing. This dataset comprises 1.7 hours of Mandarin speech and 2.9 hours of singing, alongside 3.8 hours of English speech and 5.8 hours of singing. 28 singers are recruited and paid to sing the songs and read the transcripts separately. To enrich the dataset, fine-grained F0 contours and MIDI annotations are extracted and created for detailed pitch information \cite{li2024robust}. Upon completing the training phase, we employ zero-shot inference on the speech and pitch inputs for comprehensive final evaluation.

\begin{table*}
\setlength\tabcolsep{10pt}
\centering
\setlength{\belowcaptionskip}{-0.4cm}
\begin{tabular*}{0.91\hsize}{l|cc|ccc}
\toprule
\bf Method                          & \bf LSD $\downarrow$& \bf RCA $\uparrow$  & \bf MOS-S $\uparrow$    & \bf MOS-P $\uparrow$  & \bf MOS-Q $\uparrow$  \\
\midrule
GT (vocoder)                        & 2.512               & 0.988               & 4.52$\pm$0.05           & 4.47$\pm$0.11         & 4.13$\pm$0.07         \\
\midrule[0.2pt]
\cite{parekh2020speech}             & 8.045               & 0.842               & 3.02$\pm$0.13           & 2.99$\pm$0.09         & 3.01$\pm$0.12         \\
\cite{wu2020speech}                 & 6.913               & 0.896               & 3.12$\pm$0.10           & 3.21$\pm$0.13         & 3.15$\pm$0.10         \\
AlignSTS                            & 5.519               & 0.941               & 3.45$\pm$0.05           & 3.47$\pm$0.06         & 3.41$\pm$0.04         \\
\midrule[0.2pt]
SVPT (18-layer feat)                & 5.462               & 0.956               & 3.44$\pm$0.06           & 3.48$\pm$0.09         & 3.39$\pm$0.08        \\
SVPT (Mandarin)                     & 5.066               & 0.982               & 3.61$\pm$0.05           & 3.76$\pm$0.06         & 3.69$\pm$0.05         \\
\midrule[0.2pt]
SVPT                                & 5.213               & 0.967               & 3.46$\pm$0.04           & 3.68$\pm$0.08         & 3.61$\pm$0.06         \\
\bottomrule
\end{tabular*}
\caption{\label{tab:main-results}
The evaluation results of STS systems using the English test set, except "SVPT (Mandarin)".
}
\end{table*}

\subsection{Implementation}

We build a 20-layer global Transformer and a 6-layer local Transformer as the backbone network. A 24-layer XLSR-53 pre-trained model is utilized, where the 12th layer is used for semantic extraction. For acoustic extraction, we train a SoundStream model for 16kHz audios with 8 codebooks. Both the semantic and acoustic extraction share a downsampling rate of 320. A unit-based vocoder, BigVGAN is trained to upsample and reconstruct 24k audios from acoustic tokens of 3 codebooks. Continuous random resampling with $N_r$ set to 20 is adopted. More details are listed in \autoref{sec:arch}.


\subsection{Evaluation Metrics}

The performance evaluation comprises two main aspects: objective and subjective assessments. For objective evaluation, we adhere to established practices \cite{parekh2020speech} and employ log-spectral distance (LSD) to gauge overall reconstruction quality.
For pitch regression assessment, we calculate the F0 raw chroma accuracy (RCA) using the \texttt{mir\_eval} library \cite{raffel2014mir_eval}, with a maximum pitch tolerance set at 50 cents. 

For subjective evaluation metrics, we utilize the Mean Opinion Score (MOS) to evaluate synthesis quality. Specifically, MOS-Q assesses overall quality, while MOS-P focuses on the naturalness and coherence of prosody. Additionally, we score MOS-S to measure singer similarity in terms of timbre and prosody. In the ablation study, the Comparative Mean Opinion Score is used to assess synthesis quality, denoted as CMOS-X. More details are listed in \autoref{sec:eval}.


\subsection{Baseline Models}

We compare the performance of SVPT with other STS methods: 1) GT (vocoder), where we compare our model with the audio generated by the BigVGAN vocoder from GT acoustic codes, which is the upper limit; 2) \cite{parekh2020speech}, a CNN-based STS model, reproduced and retrained using the NHSS database; 3) \cite{wu2020speech}, a GAN-based model designed to further improve synthesis quality, reproduced and retrained; 4) AlignSTS, a diffusion-based model with rhythm and pitch modal fusion, retrained. Since only English STS datasets are available, the retrained STS baselines are monolingual, and we only compare the methods using the English test set. However, to provide a complete comparison, we record the performance of our model tested on the Mandarin test set, denoted as "SVPT (Mandarin)". In addition, we compare the model where the semantic features are from the 18th layer of the wav2vec 2.0 model, denoted as "SVPT (18-layer feat)".

For the extensional experiments of SVS, we provide the results of two non-autoregressive SVS models as baselines: 1) FFT-Singer, an SVS model based on stacked feed-forward transformer blocks; and 2) DiffSinger \cite{Liu_Li_Ren_Chen_Zhao_2022}, which is based on denoising diffusion probabilistic model. Since SVPT is for zero-shot generation, we add a global timbre encoder module to each baseline by leveraging a speaker identity encoder\footnote{https://github.com/resemble-ai/Resemblyzer}. Furthermore, we utilize a HiFi-GAN vocoder \cite{kong2020hifi} for the baselines to transform the Mel-spectrograms into waveforms. 

\begin{table*}
\setlength\tabcolsep{10pt}
\centering
\setlength{\belowcaptionskip}{-0.4cm}
\begin{tabular*}{0.84\hsize}{l|cc|ccc}
\toprule
\bf Method                          & \bf LSD $\downarrow$& \bf RCA $\uparrow$  & \bf MOS-S $\uparrow$    & \bf MOS-P $\uparrow$  & \bf MOS-Q $\uparrow$  \\
\midrule
GT (HiFi-GAN)                       & 2.392               & 0.992               & 4.73$\pm$0.07           & 4.54$\pm$0.07         & 4.27$\pm$0.06         \\
GT (BigVGAN)                        & 2.364               & 0.993               & 4.72$\pm$0.04           & 4.56$\pm$0.06         & 4.30$\pm$0.04         \\
\midrule[0.2pt]
FFT-Singer                          & 4.859               & 0.942               & 3.24$\pm$0.05           & 3.37$\pm$0.04         & 3.29$\pm$0.06         \\
DiffSinger                          & 4.643               & 0.967               & 3.30$\pm$0.04           & 3.61$\pm$0.06         & 3.60$\pm$0.04         \\
\midrule[0.2pt]
SVPT                                & 4.750               & 0.971               & 3.48$\pm$0.06           & 3.57$\pm$0.07         & 3.56$\pm$0.06         \\
\bottomrule
\end{tabular*}
\caption{\label{tab:svs}
The evaluation results of SVS systems using the Mandarin test set. 
}
\end{table*}

\subsection{Main Results}

The main results are shown in \autoref{tab:main-results}. From the objective and subjective results, we can see that despite our model being trained primarily on data in Mandarin, it still outperforms the baselines on the English test set. 
The MOS-S scores of the first two baselines are particularly lacking, possibly because they do not have special designs for the extraction or reconstruction of timbre information. AlignSTS adopts a continuous global timbre extractor, while our model utilizes a discrete reference prompt and leverages the in-context learning capability. The superior performance of our model also reflects the importance of the model's scalability to data, meaning our model can be trained with more unlabeled data or even cross-lingual data. This also indicates that for spoken language models, different languages have a mutual promotional effect on each other. Another important reason is that STS tasks focus more on pronunciation patterns instead of semantic patterns. That is, STS models only map the syllables to the target rhythm and pitch frames, without regard to the semantic meanings. Therefore, as long as the two languages share partially similar phonemes, mutual promotion exists. 
This finding is also corroborated in the results of the baseline SVPT (18-layer feat), since the features from the 12th layer are more pronunciation-related and the 18th layer is more semantic-related. 

\subsection{Ablation Study}


\begin{table}[htbp]
\setlength\tabcolsep{3pt}
\centering
\setlength{\belowcaptionskip}{-0.4cm}
\begin{tabular*}{\hsize}{c|ccc}
\toprule
\bf PRR       & \bf CMOS-S  & \bf CMOS-P       & \bf CMOS-Q  \\
\midrule
discrete      & +0.02       & +0.03            & -0.01       \\
\midrule[0.2pt]
$N_r$ = 1     & -0.15       & -0.42            & -0.31       \\
$N_r$ = 5     & -0.09       & -0.18            & -0.12       \\
$N_r$ = 10    & -0.03       & -0.04            & -0.02       \\
$N_r$ = 20 (ours) & 0.00        & 0.00             & 0.00        \\
$N_r$ = 50    & 0.00       & +0.01            & +0.02           \\
\bottomrule
\end{tabular*}
\caption{\label{tab:prr}
Comparisons of different random resampling strategies. The first row is discrete RPP, and the rest are continuous pre-perturbations. 
}
\end{table}

We explored the effects of various PRR strategies, and the results are listed in \autoref{tab:prr}. We compare the results of discrete PPR and continuous PPR with different pre-perturbation numbers. Experimental results indicate that $N_r = 20$ yields satisfactory performance; any further enlargement of the semantic variant corpus results in only marginal improvements. On the contrary, insufficient variants can lead to severe overfitting. In addition, discrete PRR achieves considerable performance, corroborating our previous hypothesis that the noise introduced by the discrete interpolation actually enhances perturbations and creates a bottleneck. This effect essentially integrates a denoising challenge into the learning process. 

\begin{table}[htbp]
\setlength\tabcolsep{3pt}
\centering
\setlength{\belowcaptionskip}{-0.4cm}
\begin{tabular*}{0.95\hsize}{c|ccc}
\toprule
\bf Setting       & \bf CMOS-S  & \bf CMOS-P       & \bf CMOS-Q  \\
\midrule
w/o PRR           & -0.12       & -0.43            & -0.45       \\
w/o $F_p$         & -0.25       & -0.31            & -0.23       \\
w/o ERRP          & -0.18       & -0.13            & 0.08        \\
\bottomrule
\end{tabular*}
\caption{\label{tab:ablation}
Ablation study results. $F_p$ indicates the pitch and timbre corruption operation. ERRP stands for the expanded-range reference prompting. 
}
\end{table}

We also validate the effectiveness of the designs in information perturbation and reconstruction, as listed in \autoref{tab:ablation}. Firstly, we remove the random resampling step and directly use the raw semantic tokens. The results suggest that this common method in speech language models diminishes with singing voice data, due to its scarcity and high dynamics. Secondly, we omit the pitch and timbre perturbation step, resulting in a poorly reconstructed prosody. Finally, we follow the prior works in TTS and select two distinct windows from each singing sample to be the reference and the target output. The performance is also inferior because a smaller range of prompt sampling can lead to overfitting. 

\subsection{Zero-shot Singing Voice Synthesis}

In this section, we test the combination of SVPT and the auxiliary text-to-semantic translator with other non-autoregressive SVS methods, using the Mandarin test sets only. Initially, the auxiliary translator undergoes pre-training on the AISHELL-3 dataset. Subsequently, it is fine-tuned using a selection of annotated datasets: M4Singer, Opencpop, and internal \#1. From the results listed in \autoref{tab:svs}, we can see that SVPT possesses a considerable capability of text-based singing voice synthesis. Although SVPT does not match the overall quality of DiffSinger, it offers insights into scaling up SVS models. Moreover, our model exhibits higher timbre similarity in zero-shot inference, suggesting the important advantage of spoken language models. 

\section{Conclusion}

This paper proposed SVPT, the first STS approach boosted by a self-supervised singing voice pre-training model, which further improved singing voice synthesis. We adopted discrete-unit random resampling and pitch corruption strategies to stabilize and generalize the training of the spoken language model. With an external text-to-semantic translator, SVPT served as an effective backbone for SVS and provided an opportunity for SVS models to scale up. Experimental results revealed notable performances in both STS and SVS tasks. 

\section*{Limitations and Potential Risks}

The proposed method presents two significant limitations. Firstly, the training and inference procedure only involves fine-grained F0 contours, which may not be applicable in practice. Secondly, Utilizing language models incurs significant computational overhead, necessitating further experiments to ascertain whether such computational demands are justified by efficiency gains.

Additionally, there exists a potential for copyright infringement concerns stemming from the improper application of the proposed STS model. To counteract these issues, we plan to introduce suitable measures aimed at preventing any unlawful or unauthorized exploitation of the technology.

\section*{Acknowledgement}

This work was supported in part by the National Natural Science Foundation of China under Grant No.62222211.

\bibliography{acl_latex}

\begin{thebibliography}{51}
\expandafter\ifx\csname natexlab\endcsname\relax\def\natexlab#1{#1}\fi

\bibitem[{Agostinelli et~al.(2023)Agostinelli, Denk, Borsos, Engel, Verzetti, Caillon, Huang, Jansen, Roberts, Tagliasacchi et~al.}]{agostinelli2023musiclm}
Andrea Agostinelli, Timo~I Denk, Zal{\'a}n Borsos, Jesse Engel, Mauro Verzetti, Antoine Caillon, Qingqing Huang, Aren Jansen, Adam Roberts, Marco Tagliasacchi, et~al. 2023.
\newblock Musiclm: Generating music from text.
\newblock \emph{arXiv preprint arXiv:2301.11325}.

\bibitem[{Baevski et~al.(2020)Baevski, Zhou, Mohamed, and Auli}]{baevski2020wav2vec}
Alexei Baevski, Yuhao Zhou, Abdelrahman Mohamed, and Michael Auli. 2020.
\newblock wav2vec 2.0: A framework for self-supervised learning of speech representations.
\newblock \emph{Advances in neural information processing systems}, 33:12449--12460.

\bibitem[{Berthelot et~al.(2017)Berthelot, Schumm, and Metz}]{berthelot2017began}
David Berthelot, Thomas Schumm, and Luke Metz. 2017.
\newblock Began: Boundary equilibrium generative adversarial networks.
\newblock \emph{arXiv preprint arXiv:1703.10717}.

\bibitem[{Borsos et~al.(2023)Borsos, Marinier, Vincent, Kharitonov, Pietquin, Sharifi, Roblek, Teboul, Grangier, Tagliasacchi, and Zeghidour}]{borsos2023audiolm}
Zalán Borsos, Raphaël Marinier, Damien Vincent, Eugene Kharitonov, Olivier Pietquin, Matt Sharifi, Dominik Roblek, Olivier Teboul, David Grangier, Marco Tagliasacchi, and Neil Zeghidour. 2023.
\newblock \href {http://arxiv.org/abs/2209.03143} {Audiolm: a language modeling approach to audio generation}.

\bibitem[{Cen et~al.(2012)Cen, Dong, and Chan}]{cen2012template}
Ling Cen, Minghui Dong, and Paul Chan. 2012.
\newblock Template-based personalized singing voice synthesis.
\newblock In \emph{2012 IEEE International Conference on Acoustics, Speech and Signal Processing (ICASSP)}, pages 4509--4512. IEEE.

\bibitem[{Chan et~al.(2022)Chan, Qian, Zhang, and Hasegawa-Johnson}]{chan2022speechsplit2}
Chak~Ho Chan, Kaizhi Qian, Yang Zhang, and Mark Hasegawa-Johnson. 2022.
\newblock Speechsplit2. 0: Unsupervised speech disentanglement for voice conversion without tuning autoencoder bottlenecks.
\newblock In \emph{ICASSP 2022-2022 IEEE International Conference on Acoustics, Speech and Signal Processing (ICASSP)}, pages 6332--6336. IEEE.

\bibitem[{Chen et~al.(2020)Chen, Tan, Luan, Qin, and Liu}]{chen2020hifisinger}
Jiawei Chen, Xu~Tan, Jian Luan, Tao Qin, and Tie-Yan Liu. 2020.
\newblock Hifisinger: Towards high-fidelity neural singing voice synthesis.
\newblock \emph{arXiv preprint arXiv:2009.01776}.

\bibitem[{Choi et~al.(2021)Choi, Lee, Kim, Lee, Heo, and Lee}]{choi2021neural}
Hyeong-Seok Choi, Juheon Lee, Wansoo Kim, Jie Lee, Hoon Heo, and Kyogu Lee. 2021.
\newblock Neural analysis and synthesis: Reconstructing speech from self-supervised representations.
\newblock \emph{Advances in Neural Information Processing Systems}, 34:16251--16265.

\bibitem[{Choi et~al.(2020)Choi, Kim, Park, Yong, and Nam}]{choi2020children}
Soonbeom Choi, Wonil Kim, Saebyul Park, Sangeon Yong, and Juhan Nam. 2020.
\newblock Children’s song dataset for singing voice research.
\newblock In \emph{International Society for Music Information Retrieval Conference (ISMIR)}.

\bibitem[{Chung et~al.(2021)Chung, Zhang, Han, Chiu, Qin, Pang, and Wu}]{chung2021w2v}
Yu-An Chung, Yu~Zhang, Wei Han, Chung-Cheng Chiu, James Qin, Ruoming Pang, and Yonghui Wu. 2021.
\newblock W2v-bert: Combining contrastive learning and masked language modeling for self-supervised speech pre-training.
\newblock In \emph{2021 IEEE Automatic Speech Recognition and Understanding Workshop (ASRU)}, pages 244--250. IEEE.

\bibitem[{Conneau et~al.(2020)Conneau, Baevski, Collobert, Mohamed, and Auli}]{conneau2020unsupervised}
Alexis Conneau, Alexei Baevski, Ronan Collobert, Abdelrahman Mohamed, and Michael Auli. 2020.
\newblock Unsupervised cross-lingual representation learning for speech recognition.
\newblock \emph{arXiv preprint arXiv:2006.13979}.

\bibitem[{Copet et~al.(2024)Copet, Kreuk, Gat, Remez, Kant, Synnaeve, Adi, and D{\'e}fossez}]{copet2024simple}
Jade Copet, Felix Kreuk, Itai Gat, Tal Remez, David Kant, Gabriel Synnaeve, Yossi Adi, and Alexandre D{\'e}fossez. 2024.
\newblock Simple and controllable music generation.
\newblock \emph{Advances in Neural Information Processing Systems}, 36.

\bibitem[{Duan et~al.(2013)Duan, Fang, Li, Sim, and Wang}]{duan2013nus}
Zhiyan Duan, Haotian Fang, Bo~Li, Khe~Chai Sim, and Ye~Wang. 2013.
\newblock The nus sung and spoken lyrics corpus: A quantitative comparison of singing and speech.
\newblock In \emph{2013 Asia-Pacific Signal and Information Processing Association Annual Summit and Conference}, pages 1--9. IEEE.

\bibitem[{Défossez et~al.(2022)Défossez, Copet, Synnaeve, and Adi}]{défossez2022high}
Alexandre Défossez, Jade Copet, Gabriel Synnaeve, and Yossi Adi. 2022.
\newblock \href {http://arxiv.org/abs/2210.13438} {High fidelity neural audio compression}.

\bibitem[{Huang et~al.(2022)Huang, Jansen, Lee, Ganti, Li, and Ellis}]{huang2022mulan}
Qingqing Huang, Aren Jansen, Joonseok Lee, Ravi Ganti, Judith~Yue Li, and Daniel~PW Ellis. 2022.
\newblock Mulan: A joint embedding of music audio and natural language.
\newblock \emph{arXiv preprint arXiv:2208.12415}.

\bibitem[{Huang et~al.(2021)Huang, Chen, Ren, Liu, Cui, and Zhao}]{huang2021multi}
Rongjie Huang, Feiyang Chen, Yi~Ren, Jinglin Liu, Chenye Cui, and Zhou Zhao. 2021.
\newblock Multi-singer: Fast multi-singer singing voice vocoder with a large-scale corpus.
\newblock In \emph{Proceedings of the 29th ACM International Conference on Multimedia}, pages 3945--3954.

\bibitem[{Kharitonov et~al.(2023)Kharitonov, Vincent, Borsos, Marinier, Girgin, Pietquin, Sharifi, Tagliasacchi, and Zeghidour}]{kharitonov2023speak}
Eugene Kharitonov, Damien Vincent, Zal{\'a}n Borsos, Rapha{\"e}l Marinier, Sertan Girgin, Olivier Pietquin, Matt Sharifi, Marco Tagliasacchi, and Neil Zeghidour. 2023.
\newblock Speak, read and prompt: High-fidelity text-to-speech with minimal supervision.
\newblock \emph{arXiv preprint arXiv:2302.03540}.

\bibitem[{Kim et~al.(2021)Kim, Kong, and Son}]{kim2021vits}
Jaehyeon Kim, Jungil Kong, and Juhee Son. 2021.
\newblock Vits: Conditional variational autoencoder with adversarial learning for end-to-end text-tospeech.
\newblock In \emph{Proc. ICML}, pages 5530--5540.

\bibitem[{Kong et~al.(2020)Kong, Kim, and Bae}]{kong2020hifi}
Jungil Kong, Jaehyeon Kim, and Jaekyoung Bae. 2020.
\newblock Hifi-gan: Generative adversarial networks for efficient and high fidelity speech synthesis.
\newblock \emph{Advances in Neural Information Processing Systems}, 33:17022--17033.

\bibitem[{Lakhotia et~al.(2021)Lakhotia, Kharitonov, Hsu, Adi, Polyak, Bolte, Nguyen, Copet, Baevski, Mohamed et~al.}]{lakhotia2021generative}
Kushal Lakhotia, Eugene Kharitonov, Wei-Ning Hsu, Yossi Adi, Adam Polyak, Benjamin Bolte, Tu-Anh Nguyen, Jade Copet, Alexei Baevski, Abdelrahman Mohamed, et~al. 2021.
\newblock On generative spoken language modeling from raw audio.
\newblock \emph{Transactions of the Association for Computational Linguistics}, 9:1336--1354.

\bibitem[{Lee et~al.(2022{\natexlab{a}})Lee, Ping, Ginsburg, Catanzaro, and Yoon}]{lee2022bigvgan}
Sang-gil Lee, Wei Ping, Boris Ginsburg, Bryan Catanzaro, and Sungroh Yoon. 2022{\natexlab{a}}.
\newblock Bigvgan: A universal neural vocoder with large-scale training.
\newblock \emph{arXiv preprint arXiv:2206.04658}.

\bibitem[{Lee et~al.(2022{\natexlab{b}})Lee, Kim, Lee, Song, Hwang, and Lee}]{lee2022hierspeech}
Sang-Hoon Lee, Seung-Bin Kim, Ji-Hyun Lee, Eunwoo Song, Min-Jae Hwang, and Seong-Whan Lee. 2022{\natexlab{b}}.
\newblock Hierspeech: Bridging the gap between text and speech by hierarchical variational inference using self-supervised representations for speech synthesis.
\newblock \emph{Advances in Neural Information Processing Systems}, 35:16624--16636.

\bibitem[{Li et~al.(2023)Li, Huang, Zhang, Liu, and Zhao}]{li2023alignsts}
Ruiqi Li, Rongjie Huang, Lichao Zhang, Jinglin Liu, and Zhou Zhao. 2023.
\newblock Alignsts: Speech-to-singing conversion via cross-modal alignment.
\newblock \emph{arXiv preprint arXiv:2305.04476}.

\bibitem[{Li et~al.(2024)Li, Zhang, Wang, Hong, Huang, and Zhao}]{li2024robust}
Ruiqi Li, Yu~Zhang, Yongqi Wang, Zhiqing Hong, Rongjie Huang, and Zhou Zhao. 2024.
\newblock \href {http://arxiv.org/abs/2405.09940} {Robust singing voice transcription serves synthesis}.

\bibitem[{Liu et~al.(2022{\natexlab{a}})Liu, Li, Ren, Chen, and Zhao}]{Liu_Li_Ren_Chen_Zhao_2022}
Jinglin Liu, Chengxi Li, Yi~Ren, Feiyang Chen, and Zhou Zhao. 2022{\natexlab{a}}.
\newblock \href {https://doi.org/10.1609/aaai.v36i10.21350} {Diffsinger: Singing voice synthesis via shallow diffusion mechanism}.
\newblock \emph{Proceedings of the AAAI Conference on Artificial Intelligence}, 36(10):11020--11028.

\bibitem[{Liu et~al.(2022{\natexlab{b}})Liu, Li, Ren, Zhu, and Zhao}]{liu2022learning}
Jinglin Liu, Chengxi Li, Yi~Ren, Zhiying Zhu, and Zhou Zhao. 2022{\natexlab{b}}.
\newblock Learning the beauty in songs: Neural singing voice beautifier.
\newblock \emph{arXiv preprint arXiv:2202.13277}.

\bibitem[{Min et~al.(2021)Min, Lee, Yang, and Hwang}]{min2021meta}
Dongchan Min, Dong~Bok Lee, Eunho Yang, and Sung~Ju Hwang. 2021.
\newblock Meta-stylespeech: Multi-speaker adaptive text-to-speech generation.
\newblock In \emph{International Conference on Machine Learning}, pages 7748--7759. PMLR.

\bibitem[{Panayotov et~al.(2015)Panayotov, Chen, Povey, and Khudanpur}]{panayotov2015librispeech}
Vassil Panayotov, Guoguo Chen, Daniel Povey, and Sanjeev Khudanpur. 2015.
\newblock Librispeech: an asr corpus based on public domain audio books.
\newblock In \emph{2015 IEEE international conference on acoustics, speech and signal processing (ICASSP)}, pages 5206--5210. IEEE.

\bibitem[{Parekh et~al.(2020)Parekh, Rao, and Yang}]{parekh2020speech}
Jayneel Parekh, Preeti Rao, and Yi-Hsuan Yang. 2020.
\newblock Speech-to-singing conversion in an encoder-decoder framework.
\newblock In \emph{ICASSP 2020-2020 IEEE International Conference on Acoustics, Speech and Signal Processing (ICASSP)}, pages 261--265. IEEE.

\bibitem[{Qian et~al.(2020)Qian, Zhang, Chang, Hasegawa-Johnson, and Cox}]{qian2020unsupervised}
Kaizhi Qian, Yang Zhang, Shiyu Chang, Mark Hasegawa-Johnson, and David Cox. 2020.
\newblock Unsupervised speech decomposition via triple information bottleneck.
\newblock In \emph{International Conference on Machine Learning}, pages 7836--7846. PMLR.

\bibitem[{Raffel et~al.(2014)Raffel, McFee, Humphrey, Salamon, Nieto, Liang, Ellis, and Raffel}]{raffel2014mir_eval}
Colin Raffel, Brian McFee, Eric~J Humphrey, Justin Salamon, Oriol Nieto, Dawen Liang, Daniel~PW Ellis, and C~Colin Raffel. 2014.
\newblock mir\_eval: A transparent implementation of common mir metrics.
\newblock In \emph{In Proceedings of the 15th International Society for Music Information Retrieval Conference, ISMIR}. Citeseer.

\bibitem[{Rafii et~al.(2019)Rafii, Liutkus, St{\"o}ter, Mimilakis, and Bittner}]{MUSDB18HQ}
Zafar Rafii, Antoine Liutkus, Fabian-Robert St{\"o}ter, Stylianos~Ioannis Mimilakis, and Rachel Bittner. 2019.
\newblock \href {https://doi.org/10.5281/zenodo.3338373} {{MUSDB18-HQ} - an uncompressed version of musdb18}.

\bibitem[{Saitou et~al.(2007)Saitou, Goto, Unoki, and Akagi}]{saitou2007speech}
Takeshi Saitou, Masataka Goto, Masashi Unoki, and Masato Akagi. 2007.
\newblock Speech-to-singing synthesis: Converting speaking voices to singing voices by controlling acoustic features unique to singing voices.
\newblock In \emph{2007 IEEE Workshop on Applications of Signal Processing to Audio and Acoustics}, pages 215--218. IEEE.

\bibitem[{Saitou et~al.(2004)Saitou, Tsuji, Unoki, and Akagi}]{saitou2004analysis}
Takeshi Saitou, Naoya Tsuji, Masashi Unoki, and Masato Akagi. 2004.
\newblock Analysis of acoustic features affecting" singing-ness" and its application to singing-voice synthesis from speaking-voice.
\newblock In \emph{Eighth International Conference on Spoken Language Processing}.

\bibitem[{Sharma et~al.(2021)Sharma, Gao, Vijayan, Tian, and Li}]{sharma2021nhss}
Bidisha Sharma, Xiaoxue Gao, Karthika Vijayan, Xiaohai Tian, and Haizhou Li. 2021.
\newblock Nhss: A speech and singing parallel database.
\newblock \emph{Speech Communication}, 133:9--22.

\bibitem[{Shen et~al.(2023)Shen, Ju, Tan, Liu, Leng, He, Qin, Zhao, and Bian}]{shen2023naturalspeech}
Kai Shen, Zeqian Ju, Xu~Tan, Yanqing Liu, Yichong Leng, Lei He, Tao Qin, Sheng Zhao, and Jiang Bian. 2023.
\newblock Naturalspeech 2: Latent diffusion models are natural and zero-shot speech and singing synthesizers.
\newblock \emph{arXiv preprint arXiv:2304.09116}.

\bibitem[{Shi et~al.(2020)Shi, Bu, Xu, Zhang, and Li}]{shi2020aishell}
Yao Shi, Hui Bu, Xin Xu, Shaoji Zhang, and Ming Li. 2020.
\newblock Aishell-3: A multi-speaker mandarin tts corpus and the baselines.
\newblock \emph{arXiv preprint arXiv:2010.11567}.

\bibitem[{Singla et~al.(2022)Singla, Shah, Chen, and Shah}]{singla2022audio}
Yaman~Kumar Singla, Jui Shah, Changyou Chen, and Rajiv~Ratn Shah. 2022.
\newblock What do audio transformers hear? probing their representations for language delivery \& structure.
\newblock In \emph{2022 IEEE International Conference on Data Mining Workshops (ICDMW)}, pages 910--925. IEEE.

\bibitem[{Vijayan et~al.(2017)Vijayan, Dong, and Li}]{vijayan2017dual}
Karthika Vijayan, Minghui Dong, and Haizhou Li. 2017.
\newblock A dual alignment scheme for improved speech-to-singing voice conversion.
\newblock In \emph{2017 Asia-Pacific Signal and Information Processing Association Annual Summit and Conference (APSIPA ASC)}, pages 1547--1555. IEEE.

\bibitem[{Vijayan et~al.(2018)Vijayan, Gao, and Li}]{vijayan2018analysis}
Karthika Vijayan, Xiaoxue Gao, and Haizhou Li. 2018.
\newblock Analysis of speech and singing signals for temporal alignment.
\newblock In \emph{2018 Asia-Pacific Signal and Information Processing Association Annual Summit and Conference (APSIPA ASC)}, pages 1893--1898. IEEE.

\bibitem[{Vyas et~al.(2023)Vyas, Shi, Le, Tjandra, Wu, Guo, Zhang, Zhang, Adkins, Ngan et~al.}]{vyas2023audiobox}
Apoorv Vyas, Bowen Shi, Matthew Le, Andros Tjandra, Yi-Chiao Wu, Baishan Guo, Jiemin Zhang, Xinyue Zhang, Robert Adkins, William Ngan, et~al. 2023.
\newblock Audiobox: Unified audio generation with natural language prompts.
\newblock \emph{arXiv preprint arXiv:2312.15821}.

\bibitem[{Wang et~al.(2023)Wang, Chen, Wu, Zhang, Zhou, Liu, Chen, Liu, Wang, Li, He, Zhao, and Wei}]{wang2023neural}
Chengyi Wang, Sanyuan Chen, Yu~Wu, Ziqiang Zhang, Long Zhou, Shujie Liu, Zhuo Chen, Yanqing Liu, Huaming Wang, Jinyu Li, Lei He, Sheng Zhao, and Furu Wei. 2023.
\newblock \href {http://arxiv.org/abs/2301.02111} {Neural codec language models are zero-shot text to speech synthesizers}.

\bibitem[{Wang et~al.(2022)Wang, Wang, Zhu, Wu, Li, Xue, Zhang, Xie, and Bi}]{wang2022opencpop}
Yu~Wang, Xinsheng Wang, Pengcheng Zhu, Jie Wu, Hanzhao Li, Heyang Xue, Yongmao Zhang, Lei Xie, and Mengxiao Bi. 2022.
\newblock Opencpop: A high-quality open source chinese popular song corpus for singing voice synthesis.
\newblock \emph{arXiv preprint arXiv:2201.07429}.

\bibitem[{Wilkins et~al.(2018)Wilkins, Seetharaman, Wahl, and Pardo}]{wilkins2018vocalset}
Julia Wilkins, Prem Seetharaman, Alison Wahl, and Bryan Pardo. 2018.
\newblock Vocalset: A singing voice dataset.
\newblock In \emph{ISMIR}, pages 468--474.

\bibitem[{Wu and Yang(2020)}]{wu2020speech}
Da-Yi Wu and Yi-Hsuan Yang. 2020.
\newblock Speech-to-singing conversion based on boundary equilibrium gan.
\newblock \emph{arXiv preprint arXiv:2005.13835}.

\bibitem[{Yang et~al.(2023)Yang, Tian, Tan, Huang, Liu, Chang, Shi, Zhao, Bian, Wu, Zhao, Watanabe, and Meng}]{yang2023uniaudio}
Dongchao Yang, Jinchuan Tian, Xu~Tan, Rongjie Huang, Songxiang Liu, Xuankai Chang, Jiatong Shi, Sheng Zhao, Jiang Bian, Xixin Wu, Zhou Zhao, Shinji Watanabe, and Helen Meng. 2023.
\newblock \href {http://arxiv.org/abs/2310.00704} {Uniaudio: An audio foundation model toward universal audio generation}.

\bibitem[{Zeghidour et~al.(2021)Zeghidour, Luebs, Omran, Skoglund, and Tagliasacchi}]{zeghidour2021soundstream}
Neil Zeghidour, Alejandro Luebs, Ahmed Omran, Jan Skoglund, and Marco Tagliasacchi. 2021.
\newblock \href {http://arxiv.org/abs/2107.03312} {Soundstream: An end-to-end neural audio codec}.

\bibitem[{Zhang et~al.(2022{\natexlab{a}})Zhang, Li, Wang, Deng, Liu, Ren, He, Huang, Zhu, Chen et~al.}]{zhang2022m4singer}
Lichao Zhang, Ruiqi Li, Shoutong Wang, Liqun Deng, Jinglin Liu, Yi~Ren, Jinzheng He, Rongjie Huang, Jieming Zhu, Xiao Chen, et~al. 2022{\natexlab{a}}.
\newblock M4singer: A multi-style, multi-singer and musical score provided mandarin singing corpus.
\newblock \emph{Advances in Neural Information Processing Systems}, 35:6914--6926.

\bibitem[{Zhang et~al.(2022{\natexlab{b}})Zhang, Cong, Xue, Xie, Zhu, and Bi}]{9747664}
Yongmao Zhang, Jian Cong, Heyang Xue, Lei Xie, Pengcheng Zhu, and Mengxiao Bi. 2022{\natexlab{b}}.
\newblock \href {https://doi.org/10.1109/ICASSP43922.2022.9747664} {Visinger: Variational inference with adversarial learning for end-to-end singing voice synthesis}.
\newblock In \emph{ICASSP 2022 - 2022 IEEE International Conference on Acoustics, Speech and Signal Processing (ICASSP)}, pages 7237--7241.

\bibitem[{Zhang et~al.(2023)Zhang, Huang, Li, He, Xia, Chen, Duan, Huai, and Zhao}]{zhang2023stylesinger}
Yu~Zhang, Rongjie Huang, Ruiqi Li, JinZheng He, Yan Xia, Feiyang Chen, Xinyu Duan, Baoxing Huai, and Zhou Zhao. 2023.
\newblock Stylesinger: Style transfer for out-of-domain singing voice synthesis.
\newblock \emph{arXiv preprint arXiv:2312.10741}.

\bibitem[{Zhang et~al.(2022{\natexlab{c}})Zhang, Zheng, Li, and Lu}]{zhang2022wesinger}
Zewang Zhang, Yibin Zheng, Xinhui Li, and Li~Lu. 2022{\natexlab{c}}.
\newblock Wesinger: Data-augmented singing voice synthesis with auxiliary losses.
\newblock \emph{arXiv preprint arXiv:2203.10750}.

\end{thebibliography}

\appendix

\section{Information Perturbation} \label{sec:info}

\begin{algorithm}
\caption{Pseudo Random Resampling}\label{alg:prr}
\KwData{$\boldsymbol{x} \in \mathbb{R}^T, l_r > 0, r_{\text{max}} > 0, r_{\text{min}} > 0$}
\KwResult{resampled $\widetilde{\boldsymbol{x}}$}
$N \gets \lfloor T / l_r \rfloor$\;
$N \gets \text{MAX}(1, \text{randint}(N - 1, N + 2))$\;
$\boldsymbol{r}_{\text{src}} \gets \text{random}(N) \times (r_{\text{max}} - r_{\text{min}}) + r_{\text{min}}$\;
$\boldsymbol{r}_{\text{src}} \gets \boldsymbol{r}_{\text{src}} / (\sum \boldsymbol{r}_{\text{src}} / N)$  \Comment*[r]{Maintain the length}
$\boldsymbol{l}_{\text{src}} \gets l_r \times \boldsymbol{r}_{\text{src}}$ \Comment*[r]{Lengths of source segments}
$\text{Compensate elements in }\boldsymbol{l}_{\text{src}}\text{ so that } \sum \boldsymbol{l}_{\text{src}} = T$\;
$\boldsymbol{r}_{\text{tgt}} \gets \text{random}(N) \times (r_{\text{max}} - r_{\text{min}}) + r_{\text{min}}$\;
$\boldsymbol{l}_{\text{tgt}} \gets l_r \times \boldsymbol{r}_{\text{tgt}}$ \Comment*[r]{Lengths of target segments}
$\widetilde{\boldsymbol{x}} = \varnothing$ \Comment*[r]{Empty vector}
\For{$\text{each }i \in [0, N-1]$}{
  $t \gets \sum_{j=1}^{i-1} l_{\text{src}}^{j}$\;
  $\boldsymbol{x}_i \gets \text{the segment of }\boldsymbol{x}\text{, from }t\text{ to }t+l_{\text{src}}^i$\;
  $\boldsymbol{\widetilde{x}}_i \gets \text{interpolate}(\boldsymbol{x}_i, l_{\text{tgt}}^i)$\;
  $\widetilde{\boldsymbol{x}} \gets \text{concat}(\widetilde{\boldsymbol{x}}, \boldsymbol{\widetilde{x}}_i)$\;
}
\end{algorithm}

The pseudo random resampling algorithm is shown in \autoref{alg:prr}. $l_r$ is a hyperparameter, representing the roughly average segmentation length. $r_{\text{max}}$ and $r_{\text{min}}$ represent the maximum and minimum scaling factor. $\text{randint}(A, B)$ generates a random integer in a range $[A, B)$, while $\text{random}(N)$ generates a vector with N random float numbers in a range $[0, 1)$. $\text{interpolate}(x, l)$ means interpolate the sequence $x$ into a new sequence of length $l$. In our experiments, we set $l_r$ to 0.4 seconds, or 20 frames. $r_{\text{max}}$ and $r_{\text{min}}$ are set to 1.5 and 0.5 for a symmetrical segmentation. 


To introduce pitch and timbre perturbations, we employ a sequence of three functions designed to simultaneously alter the audio information. For the formant shifting function, $fs(\cdot)$, we uniformly sample a shifting ratio from the range $(1, 1.4)$. Following the ratio sampling, we make a random decision on whether to invert the sampled ratio.
In the case of $pr(\cdot)$, which addresses pitch perturbations, both a pitch shift ratio and a pitch range ratio are uniformly sampled from the intervals $(1, 2)$ and $(1, 1.5)$, respectively. Similar to the formant shifting process, we determine randomly whether to invert these sampled ratios.
The $peq(\cdot)$ function denotes a combination of audio filters: one low-shelving filter (HLS), one high-shelving filter (HHS), and eight peaking filters (HPeak), structured to modify the audio's equalization profile. This approach allows for nuanced adjustments to both pitch and timbre, enhancing the diversity and realism of the synthesized audio.

\section{Architecture Details} \label{sec:arch}

\begin{table}
\centering
\setlength{\belowcaptionskip}{-0.4cm}
\begin{tabular*}{0.9\hsize}{l|c|c}
\toprule
\multicolumn{2}{c|}{Hyperparameter}                            & Model    \\
\midrule
\multirow{5}*{\shortstack{Global\\Transformer}}          & Hidden Size               & 192     \\
~                                                       & Layers                    & 20     \\
~                                                       & Hidden Dim                & 1152   \\
~                                                       & Attention Heads           & 16   \\
~                                                       & FFN Dim                   & 4608   \\
\midrule[0.2pt]
\multicolumn{2}{c|}{Number of Parameters}                     & 320.1M    \\
\midrule[0.2pt]
\multirow{5}*{\shortstack{Local\\Transformer}}           & Hidden Size               & 192     \\
~                                                       & Layers                    & 6      \\
~                                                       & Hidden Dim                & 1152   \\
~                                                       & Attention Heads           & 8   \\
~                                                       & FFN Dim                   & 4608   \\
\midrule[0.2pt]
\multicolumn{2}{c|}{Number of Parameters}                     & 100.1M    \\
\multicolumn{2}{c|}{Total Number of Parameters}                     & 420.2M    \\
\bottomrule
\end{tabular*}
\caption{\label{tab:arch}
Hyperparameters of the multi-scale Transformer.
}
\end{table}

The overall configuration of the multi-scale Transformer is listed in \autoref{tab:arch}. 
We train the multi-scale Transformer on the 200-hour corpus with 6 NVIDIA V100 GPUs with a batch size of 4000 tokens for each GPU, for 6 days. We combine the singing voice datasets listed in \autoref{tab:train-data} and the mix of AISHELL-3 and LibriSpeech, a total of about 630 hours, to train the semantic and acoustic extractors. We finetune the XLSR-53 model with 2 NVIDIA V100 GPUs with a batch size of 1200k tokens for 50k steps. We train the SoundStream model from scratch with 16 NVIDIA V100 GPUs with a batch size of 1800k tokens for 1000k steps. To integrate prosody knowledge of singing voices, we finetune XLSR-53 on about 200 hours of singing voice data in 4 languages. 


\section{Evaluation Details} \label{sec:eval}

For the subjective evaluation of STS and SVS tasks, we randomly chose 20 samples from the test set for subjective analysis. 15 professional listeners were recruited to evaluate these samples. The Mean Opinion Score-Quality (MOS-Q) assessment focused on the overall quality of the synthesized singing, including aspects such as clarity and naturalness. Meanwhile, for the Mean Opinion Score-Pitch (MOS-P), evaluators listened to Ground Truth (GT) samples with instructions to focus solely on the accuracy of pitch reconstruction, overlooking the quality of the audio. 
Furthermore, we utilize the Similarity Mean Opinion Score (MOS-S) \cite{min2021meta} to evaluate the reconstruction in timbre and prosody between the synthesized and reference singing samples. Ratings across these evaluations are based on a Likert scale from 1 to 5. presented alongside 95\% confidence intervals.  During our ablation studies, the Comparative Mean Opinion Score (CMOS) is applied for a nuanced assessment of synthesis quality, with specific focuses delineated as CMOS-S, CMOS-Q, and CMOS-P. For CMOS and its variants, participants compare pairs of singing samples from different systems to indicate their preference, utilizing a scale where 0 indicates no perceptible difference, 1 indicates a minor difference, and 2 signifies a major difference. It's essential to acknowledge that all evaluators were remunerated at an hourly rate of \$8, cumulating an overall expense of approximately \$400 for their participation. They were duly notified that their input is instrumental for scientific research purposes.

\section{Additional Discussion}

\paragraph{Disentangled Semantic Tokens}

We believe the speaker identification information is removed after the perturbation and tokenization. However, we find that this information is disentangled even before the perturbation. To evaluate the semantic tokens, we perform a special experiment to train the BigVGAN using only semantic tokens as input (original extracted representations, no perturbations), instead of acoustic tokens. After convergence, we find that the generated voices have undetermined timbre characteristics. That is, the timbre will change frequently and randomly along the time axis within one sample, even cross-gender. Therefore, with the additional perturbations, we believe the timbre information is removed. 

\paragraph{Choice of $l_r$} \label{sec:lr}

$l_r$ = 0.4 is a common choice in the previous random resampling methods. For example, \citet{qian2020unsupervised} adopts 0.304s as the minimum segment length and 0.512s as the maximum (this can be found in their code). This is also a reasonable choice since, on average there are about 3.7 phonemes per second, in the annotated portion of our datasets. That is, the average phoneme duration is about 0.27 seconds, and the minimum segment length in our setting is $l_r \times r_\text{min} = 0.4 \times 0.5 = 0.2$ seconds. However, experiments show that this choice does not entail strong exclusivity, we believe any choice between 0.35-0.5 would work. For a larger $l_r$, the effect of rhythm disentanglement will be degraded, resulting in a higher possibility that the generation inherits the rhythm of the input speech sample during zero-shot inference. For a smaller $l_r$, convergence is much harder, and the generation may contain phonemes that are not present in the input speech during inference. This may be because shrinking smaller segments causes severe information loss, forcing the model to create phonemes during generation.

\paragraph{Regarding AlignSTS}

For the claim in the Introduction Section: the attention map of AlignSTS has a chance to degrade when facing long sentences, we provide an additional discussion. Ideally, the attention weights should look monotonically. However, when encountering sentences with many syllables or a rapid pace, the attention weights begin to fragment and gradually become low-rank. 

For comparison, we trained the proposed model with the datasets listed in \autoref{tab:train-data} combined with two speech corpora. The baselines (including AlignSTS) were retrained with NHSS and PopBuTFy. In the testing stage, we use the English subset of the dataset in Section \ref{sec:exp-data} for a fair comparison, since the baselines are retrained with only English datasets. We believe that the performance reduction of AlignSTS has two reasons: 1) Indeed, the test sets are different, in that the test samples used by AlignSTS are generally shorter in length, and the performance tends to drop significantly facing longer signals; 2) the results of MOS scores are heavily influenced by subjective factors. 

\end{document}